\documentclass[%
 reprint,
 amsmath,amssymb,
 aps,
prb,
]{revtex4-1}

\usepackage{graphicx}
\usepackage{dcolumn}
\usepackage{bm}

\usepackage{comment}
\usepackage{physics}
\usepackage{color}

\def\GVec#1{\mbox{\boldmath $#1$}}

\begin{document}


\title{
Topological junction states and their crystalline network in chiral-symmetric systems:
application to graphene nanoribbons
}


\author{Gen Tamaki}
\author{Takuto Kawakami}%
\author{Mikito Koshino}%
\affiliation{%
Department of Physics, Osaka University, Osaka 560-0043, Japan
}%

\date{\today}

\begin{abstract}
We develop a general theoretical framework based on $Z$-classification to count the number of topological bound states
at a junction of chiral-symmetric one-dimensional systems.
The formulation applies to general multiway junctions composed of an arbitrary number of channels and an arbitrary joint structure.
By using the formula, we calculate the zero-energy bound states in various types of two-way and three-way junctions
of semiconducting graphene nanoribbons.
We then consider periodic two-dimensional networks of graphene nanoribbons,
and show that the topological junction states form isolated energy bands inside the bulk energy gap,
which can be viewed as a two-dimensional crystal of the effective atoms.
Depending on the $Z$ number of a single junction, we have a different set of effective atomic orbitals, resulting in
various types of nanoscale metamaterials, which are often accompanied by flat bands.
The system would provide an ideal platform for quantum simulator
to emulate a strongly-interacting fermion system on various types of lattices.
\end{abstract}

\maketitle


\section{Introduction}

The topological characterization is now recognized as a fundamental tool to understand 
electronic properties of materials. One of the most striking effects caused by nontrivial topology
is the emergence of surface states at the boundary between topologically distinct materials.
\cite{fu2007topological,hasan2010colloquium,qi2011topological}
When different topological materials are periodically arranged in a superlattice, 
a serial connection of mutually coupled interface states often gives rise to novel quantum phases,
\cite{burkov2011weyl,li2014superlattice,belopolski2017novel,shibayev2019engineering,crasto2019layertronic}
providing a powerful approach to the topological band engineering.

\begin{figure}[h]
 \centering
 \includegraphics[width=65mm]{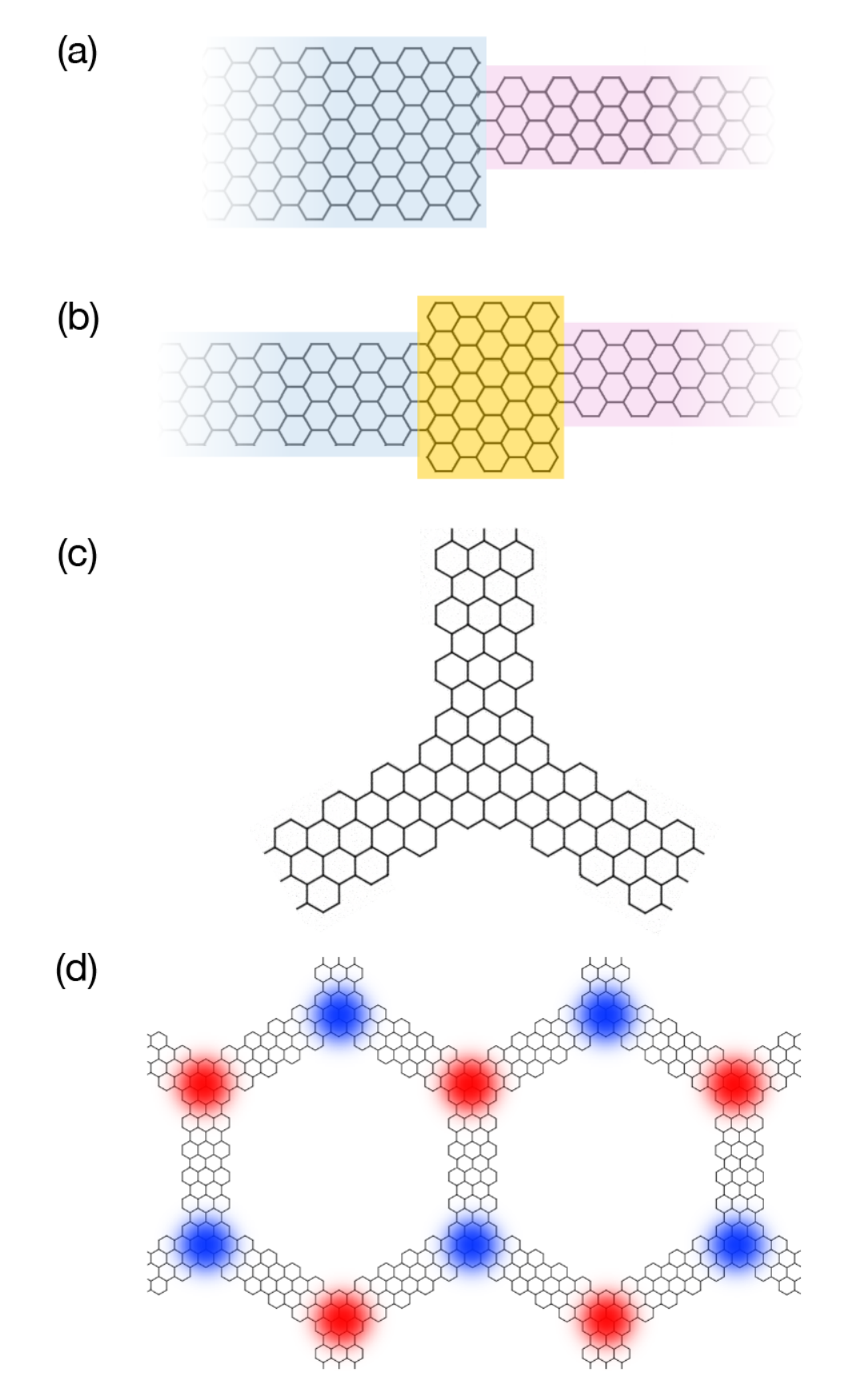}
 \caption{Examples of chiral symmetric junctions to which the developed theory is applicable.
(a) Two-way junction from a direct connection of two GNRs. 
(b) Two-way junction of two GNRs with an intermediate section.
(c) Three-way junction.
(d) GNR network as a 2D crystal of topological junction states.
 }
 \label{fig_theme}
\end{figure}

Recently, graphene nanoribbons (GNRs) \cite{nakada1996edge,wakabayashi1999electronic,son2006energy,son2006half,han2007energy,novoselov2012roadmap}
have drawn attention as one-dimensional (1D) topological materials.
It was shown that the armchair-edged GNR is characterized by $Z_2$ invariant, and
the topological bound states exist at the interface of GNRs belonging to different classes. \cite{cao2017topological,lee2018topological}
The argument was extended to the chiral symmetric materials with $Z$ classification. \cite{jiang2020topology}
It was also proposed that a 1D superlattice of GNR junction states form an interacting spin chain,
which would offer an ideal platform to study quantum spin effects in 1D.\cite{cao2017topological,jiang2020topology}
Experimentally, a recent development of nanofabrication techniques  
\cite{cai2010atomically,bennett2013bottom,chen2013tuning,narita2014bottom,cai2014graphene,chen2015molecular,
narita2015bottom,kawai2015atomically,ruffieux2016surface,wang2018bottom,kojima2019bottom,teeter2019surface,sun2020massive}
realized precise control of the atomic structure of GNR.
The topological bound states were actually observed in 1D periodically modulated GNRs. \cite{rizzo2018topological,groning2018engineering}

In this paper, we expand the idea of topological engineering in GNR to a broader class of structures
including two-dimensional (2D) networks.
We develop a general theoretical framework to estimate the number of topological bound states
at a junction of chiral-symmetric systems.
In addition to the direct connection of two GNRs considered in the previous works\cite{cao2017topological,lee2018topological,jiang2020topology} [e.g., Fig.\ \ref{fig_theme}(a)],
the formulation applies to a junction with an intermediate section [Fig.\ \ref{fig_theme}(b)],
and even to multiway junctions composed of three or more GNRs [Fig.\ \ref{fig_theme}(c)].
The argument is based on the $Z$-classification under the chiral symmetry (an approximate symmetry of graphene),
and it can be applied to any chiral-symmetric systems.

By using the formula, we design 2D crystals of topological bound states
from armchair GNR networks as in Fig.\ \ref{fig_theme}(d). 
We show that the topological states sitting on junctions form a cluster of energy bands near the charge neutral point,
which are very well approximated by an ideal nearest-neighbor tight-binding model of the effective atoms.
Depending on the $Z$ number of a single junction, 
we have a different set of effective atomic orbitals, resulting in different types of effective lattice models, which are
often accompanied by flat bands with exponentially small band widths.
The zero-energy band cluster is robust against the perturbation
since they are well separated from the bulk states by the semiconducting energy gap of armchair GNRs.
The on-site Coulomb interaction in the topological bound state is typically much greater than the band width,
suggesting that the system would provide an ideal platform for quantum simulator
to emulate a strongly-interacting fermion system on various types of lattices.
The existence of the isolated topological bands inside the bulk energy gap is a major characteristic 
that distinguishes our GNR networks from the previously studied graphene superlattices with nanoscale holes 
\cite{yu2008collective,pedersen2008graphene,liu2009band,sinitskii2010patterning,bai2010graphene,liang2010formation,cui2011magic,gunst2011thermoelectric,yang2011inducing,petersen2011clar,oswald2012energy,dvorak2013bandgap,trolle2013large,power2014electronic,chen2018nanoperforated,kariyado2018counterpropagating}.
Recently, an isolated narrow band was predicted in a phenalenyl-phenyl honeycomb network, \cite{maruyama2016coexistence}
which is a type of hydrocarbon network system \cite{maruyama2016coexistence,sorimachi2017magnetic,fujii2018electronic,fujii2019three},
and it is interpreted as a topologically-nontrivial case in our approach.

The paper is organized as follows.
In Sec.\ \ref{sec_general}, we introduce the general argument to count the topological junction states
in chiral symmetric systems. In Sec.\ \ref{sec_GNR},  we apply the theory to 
various two-way and three-way junctions of armchair GNRs.
In Sec.\ \ref{sec_2D_network}, we consider 2D honeycomb networks of GNRs,
and study the energy band of the topological bound states and its effective lattice model.
The brief conclusion is given in Sec.\ \ref{sec_concl}.

\section{General theory for topological bound states}
\label{sec_general}

\begin{figure}[htbp]
 \centering
 \includegraphics[width=90mm]{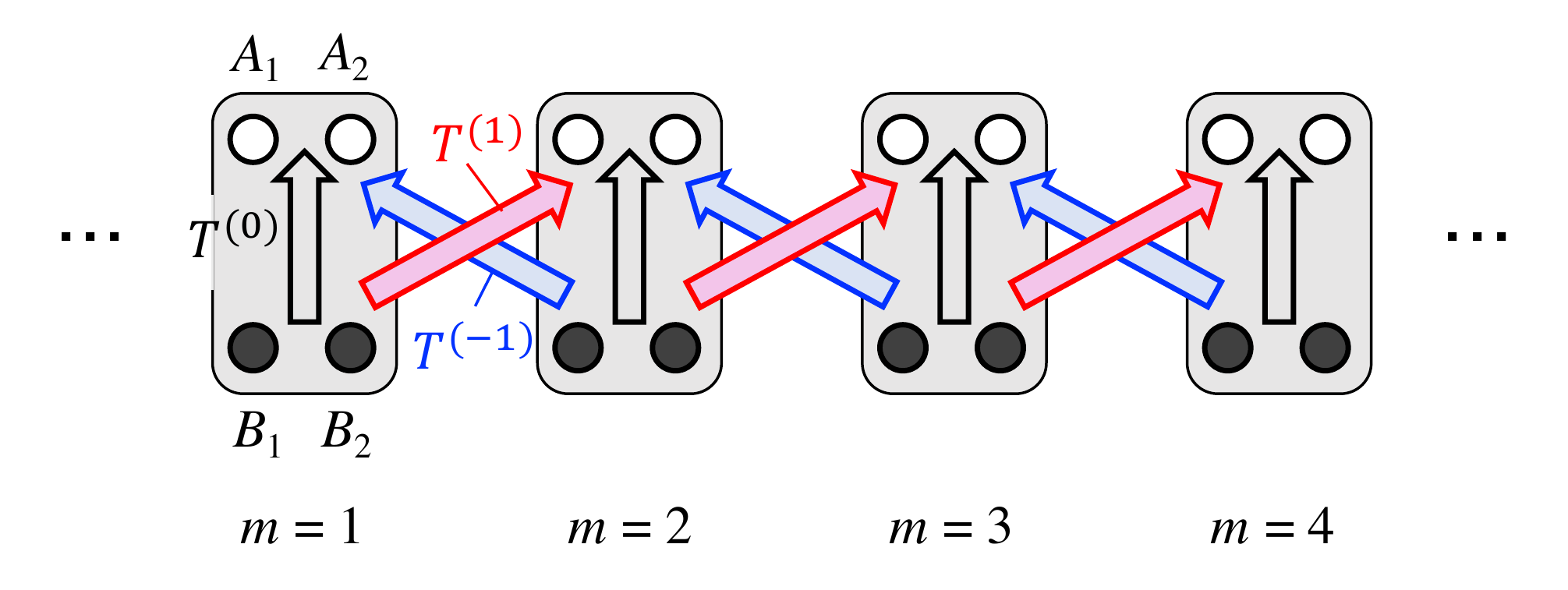}
 \caption{
1D periodic lattice with the chiral symmetry.
A unit cell contains $N$ sites of the A sublattice (open circles; $N=2$ in this figure), 
and  $N$ sites of the B sublattice (filled circles), and the 
coupling occurs only between the A sublattice and the B sublattice.
In the figure, only the intracell hopping $T^{(0)}$
and the nearest neighbor coupling  $T^{(\pm1)}$ are shown.
}
 \label{fig:chiral-ribbons}
\end{figure}

\begin{figure}
 \centering
 \includegraphics[width=80mm]{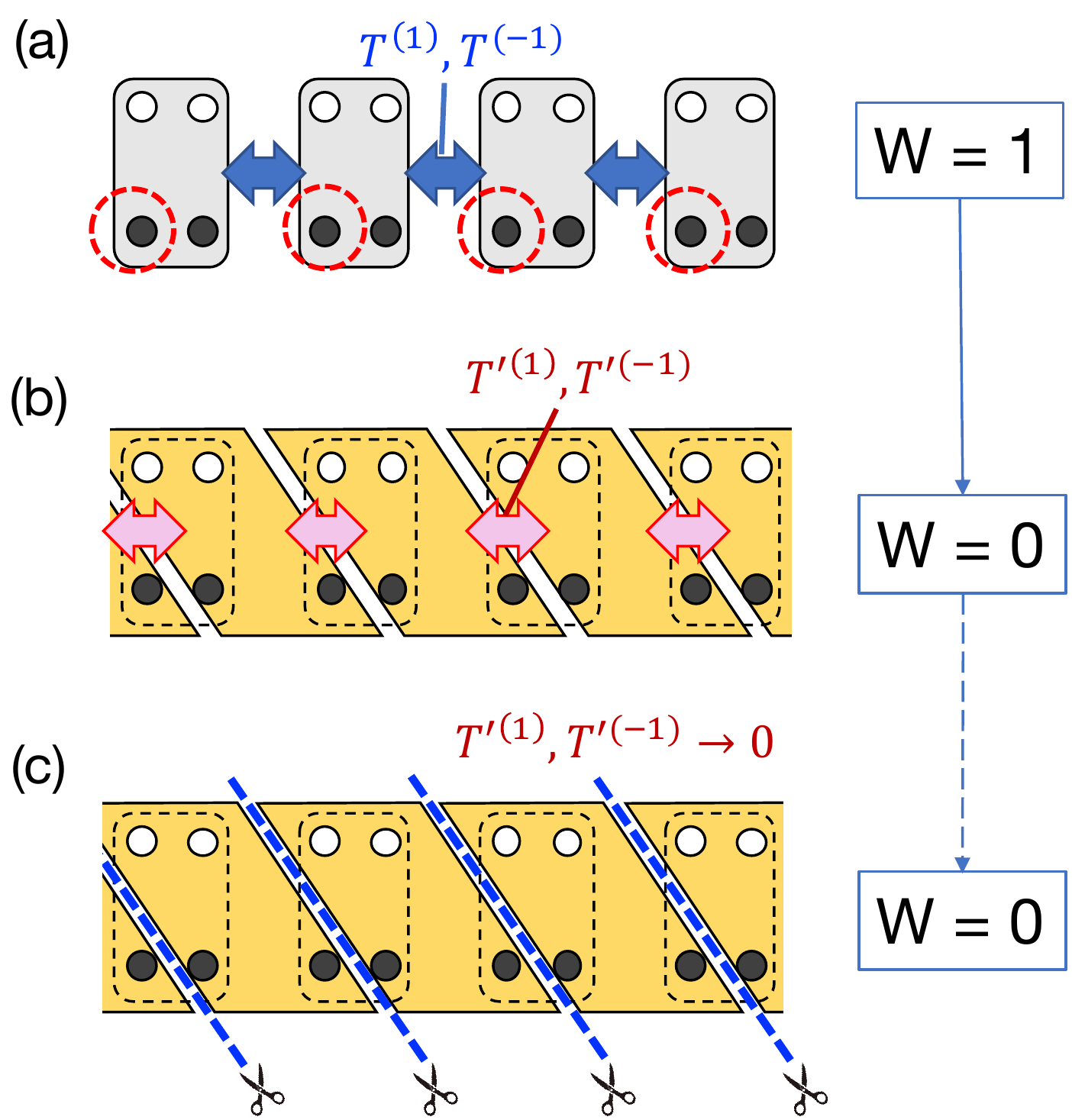}
 \caption{
 (a)(b) Modifying the unit cell by moving a $B$ site to the left.
(c) Continuously reducing all the intercell couplings $T^{(l\neq 0)}$ to zero.
 }
 \label{fig_change_W}
\end{figure}

We introduce a method to calculate the number of the topological bound states in a junction of chiral-symmetric 1D crystal.
Let us consider a 1D periodic lattice with the chiral symmetry as shown in Fig.\ \ref{fig:chiral-ribbons},
where the unit cell contains $N$ sites of the A sublattice $(A_1, A_2, \cdots, A_N)$, 
and  $N$ sites of the B sublattice $(B_1, B_2, \cdots, B_N)$.
The chiral symmetry (sublattice symmetry)
allows the couplings only between the A sublattice and the B sublattice, but not between A and A, or B and B.
The Hamiltonian is then written as
\begin{equation}
\mathcal{H}=\sum_{m,l=-\infty}^\infty\sum_{i,j=1}^N
\left(T_{ij}^{(l)} a^{(m+l)\dagger}_i  b^{(m)}_j + {\rm h.c.}\right),
\label{eq_H}
\end{equation}
where $a^{(m)\dagger}_i$ and $b^{(m)\dagger}_i\, (i=1,2,\cdots,N)$ are the creation operators for an electron 
at $A_i$ and $B_i$ in $m$-th unit cell, respectively.
The $T^{(l)}$ is the $N\times N$ matrix to describe the hopping from the B sublattice in $m$-th cell
to the A sublattice in $(m+l)$-th cell.
We assume that $T^{(l)}$ exponentially decays in increasing $|l|$, as naturally expected in real systems.
In Fig.\ \ref{fig_change_W}, only the nearest neighbor coupling  $T^{(\pm 1)}$ are shown for the illustrative purpose,
while the following argument is also valid when the further hoppings $T^{(l)} \, (|l|\geq 2)$ exist.
The simplest example of Eq.\ (\ref{eq_H}) is the Su-Schrieffer-Heeger model, \cite{su1979solitons,su1980soliton}
where $N=1$ and only $T^{(0)}$ and $T^{(1)}$ are nonzero.

The Hamiltonian Eq.\ (\ref{eq_H}) can be written in $k$-space representation as,
\begin{equation}
\mathcal{H}=\int_{-\pi}^\pi dk \Psi_k^\dagger H(k)\Psi_k,
\end{equation}
where
\begin{equation}
H(k)=\mqty(0&D(k)\\D(k)^\dagger&0),
\end{equation}
\begin{equation}
D(k)=\sum_{l=-\infty}^\infty e^{-ikl}T^{(l)},
\end{equation}
and 
\begin{equation}
\Psi_k^\dagger=\frac{1}{\sqrt{2\pi}}\sum_{m=-\infty}^\infty e^{ikm}(\vb{a}^{(m)\dagger}, \vb{b}^{(m)\dagger}).
\end{equation}
The energy band is given by the eigenvalues of $\pm D(k)D(k)^\dagger$.

If $\det D(k) \neq 0$ for any of $k$, we have a band gap at $E=0$,
and then we can define the winding number by 
\begin{equation}
W=\frac{1}{2\pi i}\int_{-\pi}^{\pi} \dd k\dv{k}\ln\det D(k).
\label{eq_W}
\end{equation}
The $W$ is always an integer and it depends on the definition of the unit cell;
it can be changed one by one by transferring a site from cell to cell,
without changing the actual physical system.
For instance, let us consider a process to move $B_j$ of a certain $j$
to the left neighboring cell, i.e., $B_j$ of $m$-th cell is renumbered to $B_j$ of $(m-1)$-th cell,
as illustrated in Figs.\ \ref{fig_change_W} (a) and (b),
Then the matrix $D(k)$ changes to $D'(k)$ where
all the elements in the $j$-th column are multiplied by $e^{-ik}$
while all the other elements remain unchanged.
As a result, we have $\det D'(k) = e^{-ik} \det D(k)$,
and then the winding number $W$ changes to $W'=W-1$ according to Eq.\ (\ref{eq_W}).
If we move an A-site to the left instead, $W$ changes to $W'=W+1$.

By repeating the process, we can always take a certain unit cell with $W=0$,
regardless of $W$ in the initial Hamiltonian.
Under this choice of the unit cell, we can continuously kill all the intercell matrices ($T^{(l)}$ for $l \neq 0$)
without closing the gap [Fig.\ \ref{fig_change_W}(c)], because $W$ remains 0 
and there is no topological phase transition during the process.
Here note that $W$ must be zero when the intercell matrices are killed,
because $D(k)$ does not depend on $k$, and then obviously $W=0$ in Eq.\ (\ref{eq_W}).
To conclude, we can continuously change the original system of Eq.\ (\ref{eq_H}) to an array of disconnected islands, 
without closing the energy gap. 

\begin{figure}
 \centering
 \includegraphics[width=85mm]{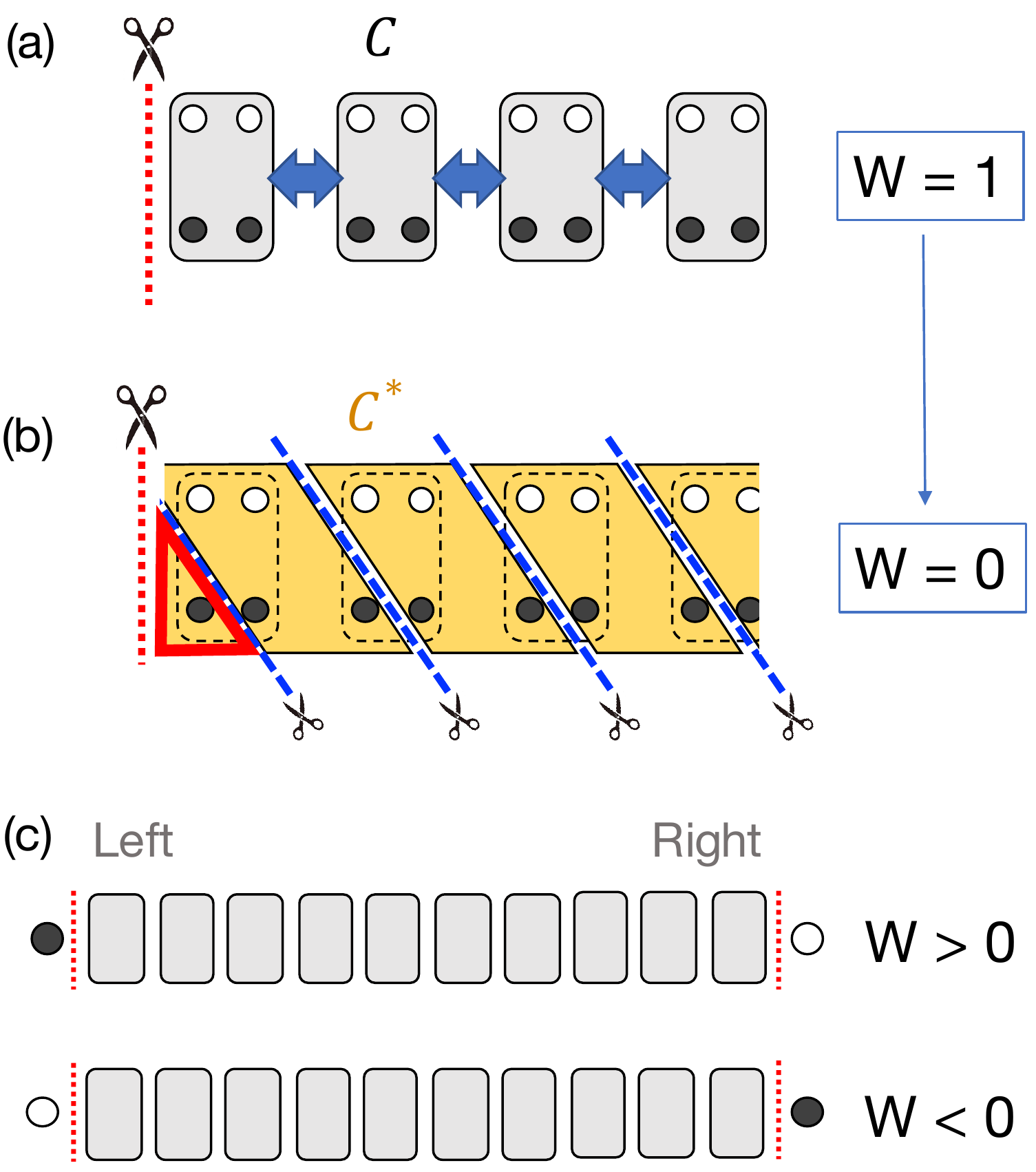}
 \caption{
(a) Semi-infinite 1D system terminated at a left boundary of the unit cell $C$.
(b) The same system as (a), where the unit cell is changed to $C^*$ with $W=0$, and then the inter-cell coupling is switched off.
(c) Relationship between the winding number $W$ and the left and right edge states in a chiral-symmetric ribbon.
 }
 \label{fig_cut}
\end{figure}

Now we introduce an edge to the system.
The number of the zero-energy edge states depends on how the system is terminated.
Here we define $C$ as an initial choice of the unit cell,
and cut the system at a boundary of $C$ as illustrated in Fig.\ \ref{fig_cut}(a).
In a chiral-symmetric system,  the existence of the zero-energy edge states is precisely correlated with the nonzero winding number. \cite{ryu2002topological}
Let $W$ be the winding number for the unit cell $C$.
It is known that \cite{prodan2016bulk} the number of the edge states at each end is equal to $|W|$,
and when $W>0(<0)$, the left and right edge states have wave amplitudes exclusively on the B and A (A and B) sublattices, respectively
[Fig.\ \ref{fig_cut}(c)].
Here the left and right edges are defined as the ends on the negative and positive $m$ sides, respectively.

This statement can be easily proved by using the continuous deformation argued above.
Let us consider the left edge of a semi-infinite system with the unit cell $C$ as in Fig.\ \ref{fig_cut}(a).
As argued, we can take a different unit cell $C^*$ which gives zero winding number.
If $W$ is positive in the original unit cell $C$, for instance, we move $W$ sites of the B sublattice from right to left to get $C^*$
[Fig.\ \ref{fig_cut} (b)].
In the terminated system, we are left with an incomplete portion [triangular part in Fig.\ \ref{fig_cut}(b)]
sandwiched by the end of the system (red dashed line) and the boundary of $C^*$ (blue dashed line).
From the definition, this end sector consists of $W$ sites of the B sublattice, which are exactly the sites transferred in changing the unit cell 
from $C$ to $C^*$.
Now we continuously switch off the intercell matrices between the boundaries of $C^*$,
to make the end sector an isolated island. Generally, a finite-sized system with the chiral symmetry
has $|N_A-N_B|$ zero-energy modes, where $N_A$ and $N_B$ are the number of sites of the A and B sublattices,
respectively.  \cite{sutherland1986localization}
These modes have the wave amplitudes on the A(B) sublattice when $N_A-N_B >0 \, (< 0)$.
In the present case, the end portion has $N_A=0$ and $N_B = W$,
and therefore it gives $W$ zero-energy modes on the B sublattice.
When we switch on the intercell matrices back to the original value,
the $W$ zero-energy modes stay at the zero energy
and remain localized near the edge with an exponential decay into the bulk. 
This is because the bulk Hamiltonian remains gapped in this process
(since the winding number does not change),
 so the amplitude of the zero-energy mode must vanish far away from the edge.
For $W<0$, we move $A$ sites instead of $B$ to find $|W|$ zero-energy edge modes on the A sublattice.
The A and B sublattices are just interchanged for the right edge.
It is concisely summarized as the following statement: 
The sublattice difference $\Delta N = N_A - N_B$ 
is given by $\mp W$ for the left and right edges, respectively.

\begin{figure}
 \centering
 \includegraphics[width=85mm]{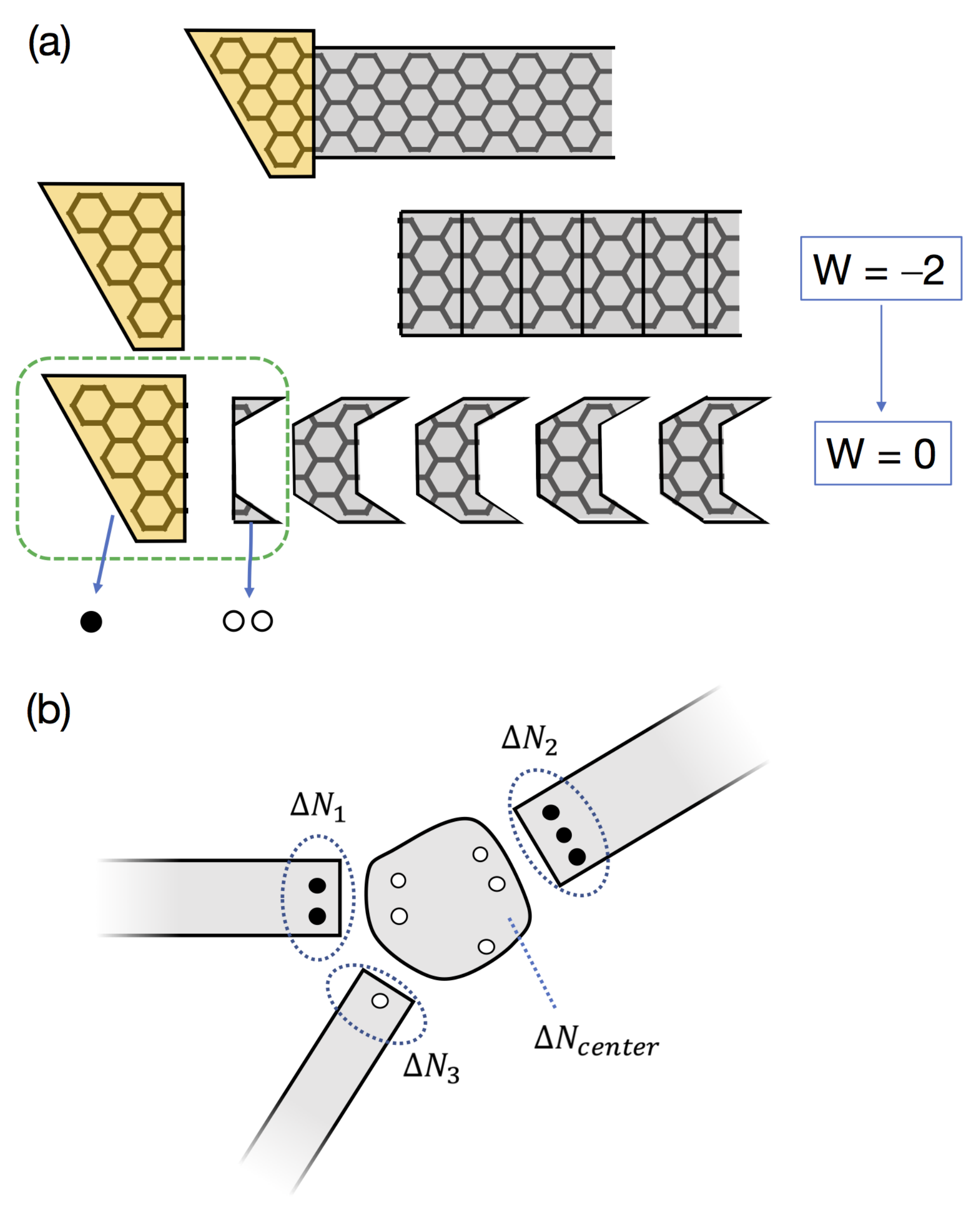}
 \caption{
(a) Example of a semi-infinite system with irregular terminal structure.
We can count the number of zero-energy edge states by 
changing the unit cell such that $W=0$, and count the sublattice difference $\Delta N = N_A - N_B$ 
in the remaining end portion (green dashed line).  
(b) A junction of two or more semi-infinite ribbons.
The total number of zero-energy states localized at the junction is given by
the summation of the end states of the ribbon and the sublattice difference in the central part. 
 }
 \label{fig_junction}
\end{figure}

The argument also applies to a system having an irregular terminal structure as in Fig.\ \ref{fig_junction}(a),
which cannot be regarded as a part of the periodic structure. 
In this case, we separate the system into the periodic semi-infinite part (gray) and the irregular terminal part (yellow).
For the periodic part, we perform the operation described above;
change the unit cell such that $W=0$.
As argued, the sublattice difference in the incomplete left end portion (gray triangles)
is given by $\Delta N = - W$, where $W$ is the winding number of the original unit cell.
For the terminal part (yellow trapezoid), we find $\Delta N_{\rm term}$ by just counting the number of A sites and B sites
of the island. Now by switching off the intercell coupling in the periodic part,
we have a combined terminal island (green dashed line)
with the sublattice difference $\Delta N_{\rm tot} =\Delta N + \Delta N_{\rm term}$.
The total number of the zero-energy edge modes is given by  $\Delta N_{\rm tot}$,
where the sign represents the sublattice A and B.

We can also consider the junction of two or more semi-infinite ribbons as illustrated in Fig.\ \ref{fig_junction}(b).
Again, we separate the system into the periodic semi-infinite ribbons ($i=1,2,\cdots$) and the central junction part.
Then find $\Delta N_i$ for the end of the ribbon $i$ from its winding number, and also $\Delta N_{\rm center}$ for the central part.
Finally, the number of the zero-energy edge modes localized at the junction 
is given by $\Delta N_{\rm tot} = \Delta N_{\rm center} + \sum_i \Delta N_i$.

\section{Graphene nanoribbon junctions}
\label{sec_GNR}

\begin{figure}
 \centering
 \includegraphics[width=55mm]{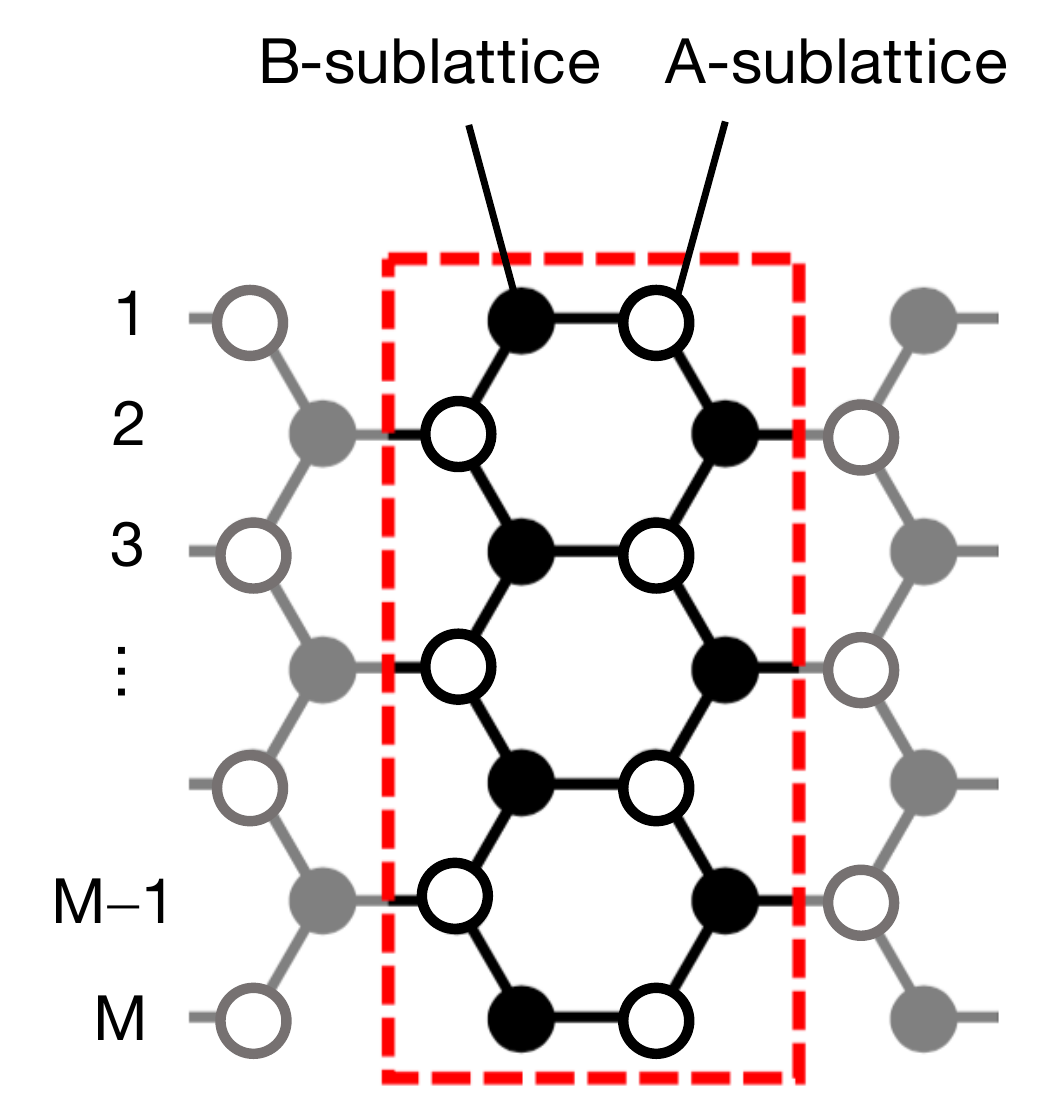}
  \caption{Atomic structure of the armchair GNR.
A unit cell (dashed square) consists of $M$ atoms of the A sublattice (open circles)
$M$ atoms of the B sublattice (solid circles).
}
 \label{fig_armchair_GNR_unit}
\end{figure}

We consider the zero-energy bound states of various GNR junctions using the argument in the previous section.
For the ribbon part, we take armchair GNRs to avoid a complication arising from the zigzag edge states at the sides of the ribbon. \cite{wakabayashi1999electronic}
We define the atomic structure of the armchair GNR as in Fig.\ \ref{fig_armchair_GNR_unit},
where a unit cell (dashed square) consists of $M$ atoms of the A sublattice (open circles)
$M$ atoms of the B sublattice (solid circles).
For the electronic Hamiltonian, we consider a single $p_z$ orbital tight-binding model on honeycomb lattice
with only the nearest neighbor hopping between A site and B site, to make the system chiral symmetric. 
We can show that the band structure of an infinite ribbon is semiconducting when $M=6k-2,6k,6k+1,6k+3$ ($k$: integer),
where the winding number is given by $W=-k$. We exclude the metallic cases, $M=6k-1,6k+2$.

\subsection{Two-ribbon junctions}

\begin{figure}
 \centering
 \includegraphics[width=75mm]{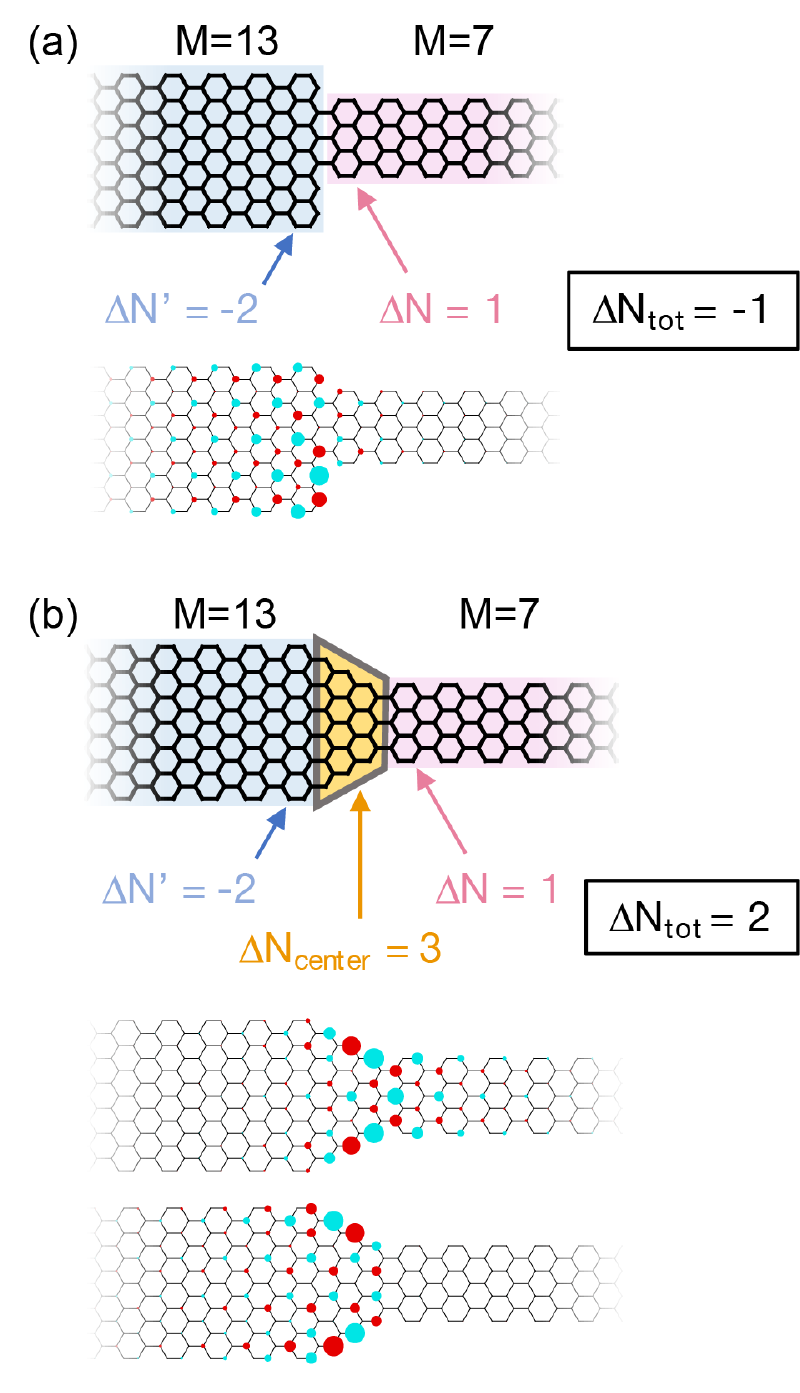}
  \caption{
 (a) Direct connection of $M=7$ ribbon and $M=13$ ribbon 
and (b) a junction of the same ribbons with an extra middle section.
The lower figures of each panel plot all the zero-energy wave functions of the system,
where the area of the circle represents the squared wave amplitude, and red and cyan correspond to positive and negative sign.
  }
 \label{fig_GNR_2-junctions}
\end{figure}

First, we consider a direct connection of $M=7$ ribbon and $M=13$ ribbon as in Fig.\ \ref{fig_GNR_2-junctions}(a).
Similar situations were also considered in the previous works.\cite{cao2017topological,lee2018topological,jiang2020topology} 
The winding numbers of  $M=7$ and 13 are given by $W=-1$ and $-2$, respectively.
According to the argument above, the left edge of the $M=7$ ribbon has $\Delta N =  1$, 
and the right edge of the $M=13$ ribbon has $\Delta N' =  -2$.
When connecting the two ends, the total sublattice difference becomes $\Delta N_{\rm tot} = \Delta N +\Delta N' = -1$,
i.e., we have a single zero-energy localized mode on the B sublattice.

We can also consider a junction as shown in Fig.\ \ref{fig_GNR_2-junctions}(b), 
which consists of the same ribbons but with an extra intermediate section.
In this example, the middle part has 15 and 12 sites in the A and B sublattices, respectively,
giving $\Delta N_{\rm center} =  3$. The total sublattice difference becomes 
$\Delta N_{\rm tot} = \Delta N +\Delta N' +\Delta N_{\rm center} = 2$, so that we have two zero-energy modes on the A sublattice.
In Fig.\ \ref{fig_GNR_2-junctions},
the lower figures of each panel show the actual zero-energy wave functions for the two cases.
There are $|\Delta N_{\rm tot}|$ zero-energy modes as expected,
and the wave functions reside on the A(B) sublattice when  $\Delta N_{\rm tot}$ is positive (negative).
In the practical calculation, we truncated the ribbons far way (at 15 unit cells from the joint part)
with armchair edges not to have zigzag edge modes.

\begin{figure*}
 \centering
 \includegraphics[width=160mm]{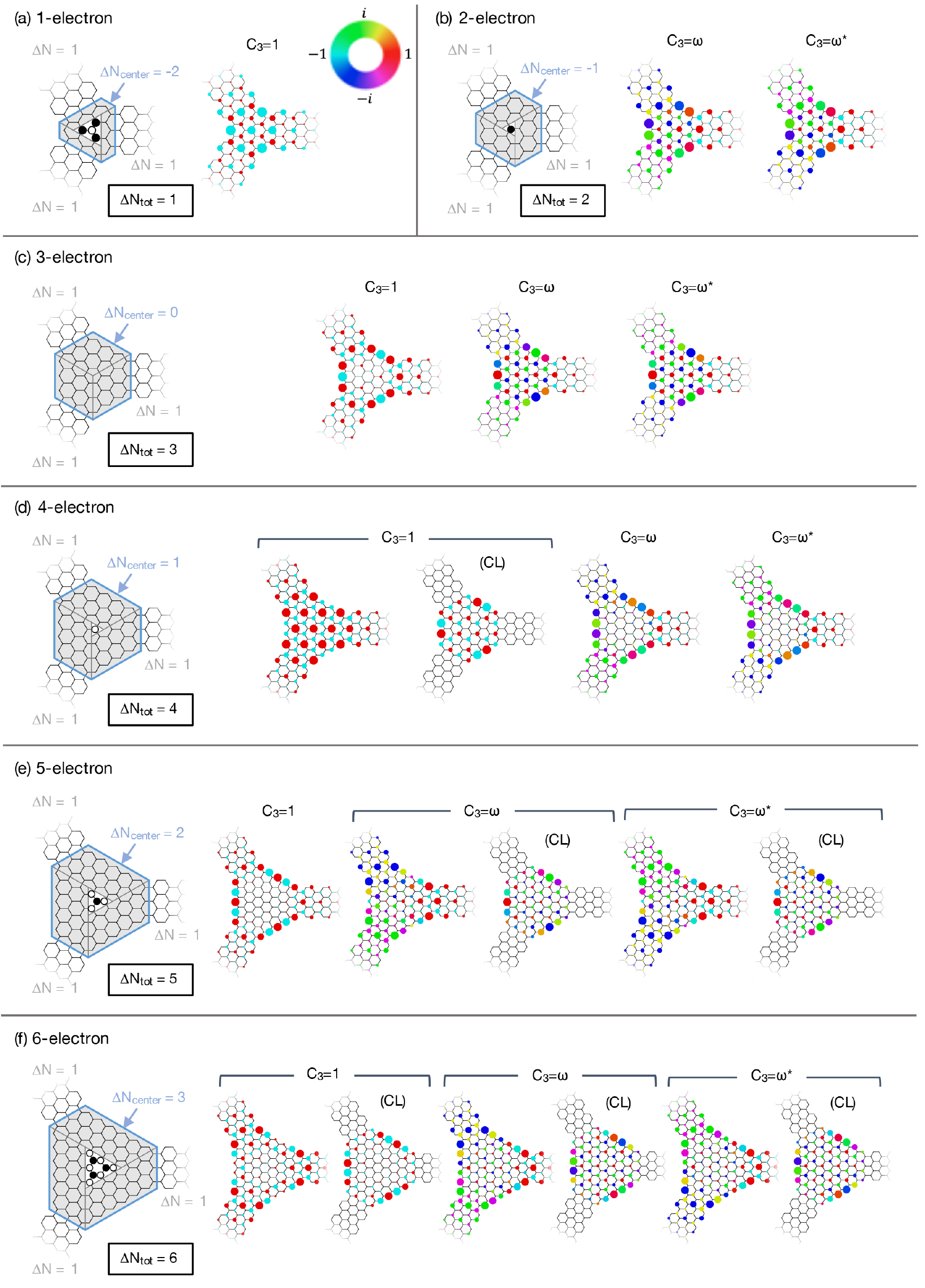}
  \caption{
Three-way junctions with from $\Delta N_{\rm tot}=1$  to 6.
In the left most panel,
a blue line defines the central section and three shaded parallelograms inside represent the regions where the numbers of A and B sites are equal,
so that $\Delta N_{\rm center}$ can be found by counting the numbers of sites in the unshaded triangular part at the center. 
The right figures plot all the zero-energy wave functions,
where the area of the circle represents the squared amplitude and the color indicates the complex phase shown in the top.
``CL'' represents the completely-localized states (see the text).
}
 \label{fig_GNR_3-junctions}
\end{figure*}

\subsection{Three-ribbon junctions}

Now let us consider triple junctions as illustrated in Fig.\ \ref{fig_GNR_3-junctions}.
Here we connect three semi-infinite $M=7$ ribbons to the central section with various different structures, which are
indicated by blue hexagons in the leftmost panels.
The shaded parallelograms in the central section represent the region where the numbers of A and B sites are equal,
so that $\Delta N_{\rm center}$ can be found from the numbers of sites 
in the remaining triangular part at the center. 
Noting the end of each $M=7$ ribbon gives $\Delta N =  1$,
the total sublattice difference of the junction is $\Delta N_{\rm tot} = \Delta N_{\rm center} + 3$.
From Fig.\ \ref{fig_GNR_3-junctions}(a) to \ref{fig_GNR_3-junctions}(f), the central part is enlarged such that
$\Delta N_{\rm center}$ increases from $-2$ to 3, so $\Delta N_{\rm tot}$ increases from 1 to 6.

Since the three-ribbon junctions considered here have $C_3$ (120$^\circ$) rotational symmetry,
the zero-energy modes can be classified by the eigenvalues of $C_3 = 1, \omega, \omega^*$,
where $\omega = \exp(2\pi i/3)$.
Such a consideration will be useful to consider the band structure of junction networks in the next section.
The sublattice difference $\Delta N_{\rm tot}$
can be divided into three sectors as $\Delta N^{(C_3=1)},\Delta N^{(C_3=\omega)},\Delta N^{(C_3=\omega^*)}$,
which are independent topological invariants.  \cite{koshino2014topological}
Table \ref{table1} shows $\Delta N^{(C_3)}$ as a function of $\Delta N_{\rm tot}$.
For $\Delta N_{\rm tot} = 4$, for instance, the numbers $(2, 1, 1)$ 
indicate that there are two zero-energy modes of $C_3 = 1$, 
and a single zero-energy mode for each of $C_3 = \omega, \omega^*$,
all in the A sublattice.

Table \ref{table1} can be obtained by the following symmetry argument without considering a specific lattice structure.
In a $C_3$ symmetric lattice, any atomic site is either 
a member of triplets (i.e., three sites located 120$^\circ$ apart) 
or the central site at the rotation origin.
If we let $\psi_1, \psi_2, \psi_3$ be three carbon-$p_z$ orbitals in a triplet (ordered in anticlockwise direction),
then $\psi_1+\psi_2+\psi_3$, $\psi_1+\omega^* \psi_2+ \omega \psi_3$ and $\psi_1+ \omega \psi_2+ \omega^*\psi_3$
are eigenstates of 120$^\circ$ rotation with $C_3=1, \omega, \omega^*$, respectively.
If there are $n_A$ and $n_B$ triplets in the A and B sublattices, respectively,
we have the sublattice difference of $n_A-n_B$ in each sector of $C_3 = 1, \omega, \omega^*$.
On top of that,
the central site, if any, always belongs to $C_3 = 1$ as it is invariant in the rotation,
so that the $C_3 = 1$ sector adds an extra sublattice difference of $1,-1$ and $0$,
when the rotational center is at A, B, and void (the center of the hexagon), respectively.
Therefore, the total sublattice difference is $\Delta N_{\rm tot} = 3(n_A-n_B)+1$, $3(n_A-n_B)-1$ and $3(n_A-n_B)$,
respectively. Table \ref{table1} is obtained by replacing $n_A-n_B$ with $n$.

\begin{table}
$
\begin{array}{| l | c | c | c | c | c | c | c | c | c | c | c |}
\hline
\Delta N_{\rm tot}  & 0 &  1 & 2 & 3  & 4 & 5 & 6 & 7 & 8 &  3n & 3n\pm 1 \\ 
\hline
C_3=1                  & 0 & 1 & 0 & 1 & 2 & 1 & 2 & 3 & 2 &  n & n\pm 1 \\ 
C_3 = \omega      & 0 & 0 & 1 & 1 & 1 & 2 & 2 & 2 & 3 &  n   & n  \\ 
C_3 =\omega^*    & 0 & 0 & 1 & 1 & 1 & 2 & 2 & 2 & 3 &  n   & n \\ 
\hline
\end{array}
$
\caption{
Distribution of the sublattice difference to three $C_3$ sectors,
as a function of $\Delta N_{\rm tot}$.
}
\label{table1}
\end{table}

The right figures in Fig.\ \ref{fig_GNR_3-junctions} plot
the actual wave functions of the zero-energy modes in each junction.
We see that the set of states obey the $C_3$ classification in Table \ref{table1}.
Generally, the zero-energy junction modes are given by hybridization
of the zero-energy modes of the central island and the edge modes of the ribbons.
In the systems of $\Delta N_{\rm tot} \geq 4$, however,
we can see that  some of the zero energy states are perfectly confined to the central island with no penetration to the ribbons,
as indicated by ``CL'' (completely localized) in Fig.\ \ref{fig_GNR_3-junctions}.
The completely-localized mode is a feature of the present nearest-neighbor hopping model.
It occurs when a zero-energy state of the central island (before connected to the ribbons)
is localized to the A(B) sublattice while the outermost sites of the island are the B(A) sublattice.
It remains an exact eigenstate of the Hamiltonian even when the outermost sites are connected to the external sites,
because the state has zero amplitudes for the outermost sites.

\begin{figure}
 \centering
 \includegraphics[width=80mm]{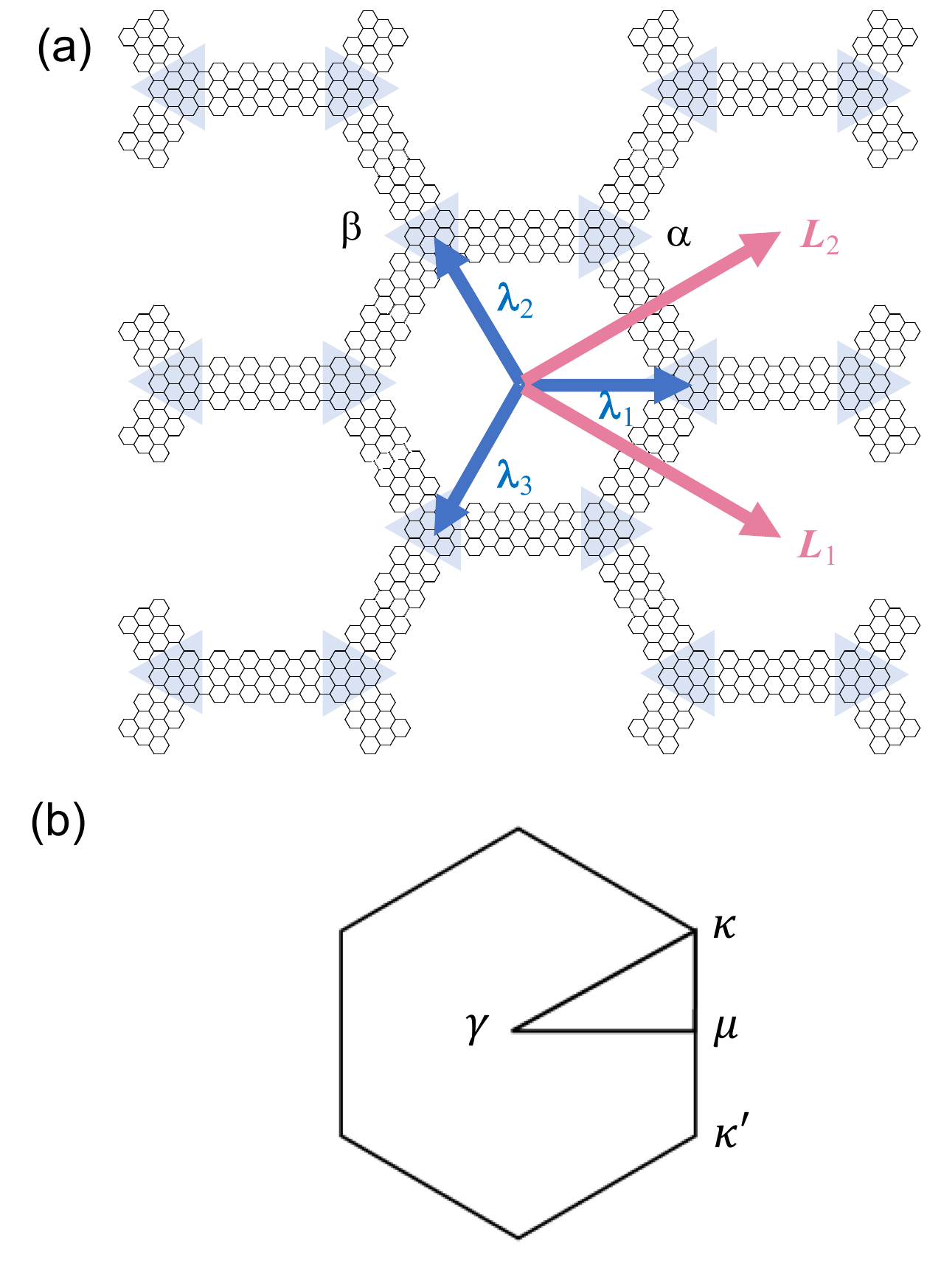}
  \caption{
(a) Example of 2D network of GNR,
where the 1-electron junctions of Fig.\ \ref{fig_GNR_3-junctions}(a)
are connected by $M=7$ ribbons of 4 unit cell long.
(b) The corresponding Brillouin zone and symmetric points.
  }
 \label{fig_GNR_network}
\end{figure}

\section{2D network of topological zero modes in GNR}
\label{sec_2D_network}

By arranging the chiral-symmetric junctions in a periodic manner,
we can have a 2D crystal of the topological bound states.
Here we consider the honeycomb lattice of armchair GNRs 
by connecting the Y-shaped junctions argued in the previous sections.
Figure \ref{fig_GNR_network}(a) shows an example,
where the 1-electron junctions of Fig.\ \ref{fig_GNR_3-junctions}(a)
are connected by $M=7$ ribbons of 4 unit cell long.
We define $\alpha$ and $\beta$ as the sublattices in the super structure as in the figure.
We also define $\mathbf{L}_1$ and $\mathbf{L}_2$ as the primitive lattice vectors of the supercell 
and $\GVec{\lambda}_l\, (l=1,2,3)$ as vectors from the center of the hollow to three inequivalent $\beta$ sites.
The corresponding Brillouin zone is shown in Fig.\ \ref{fig_GNR_network}(b).

\begin{figure*}
 \centering
 \includegraphics[width=170mm]{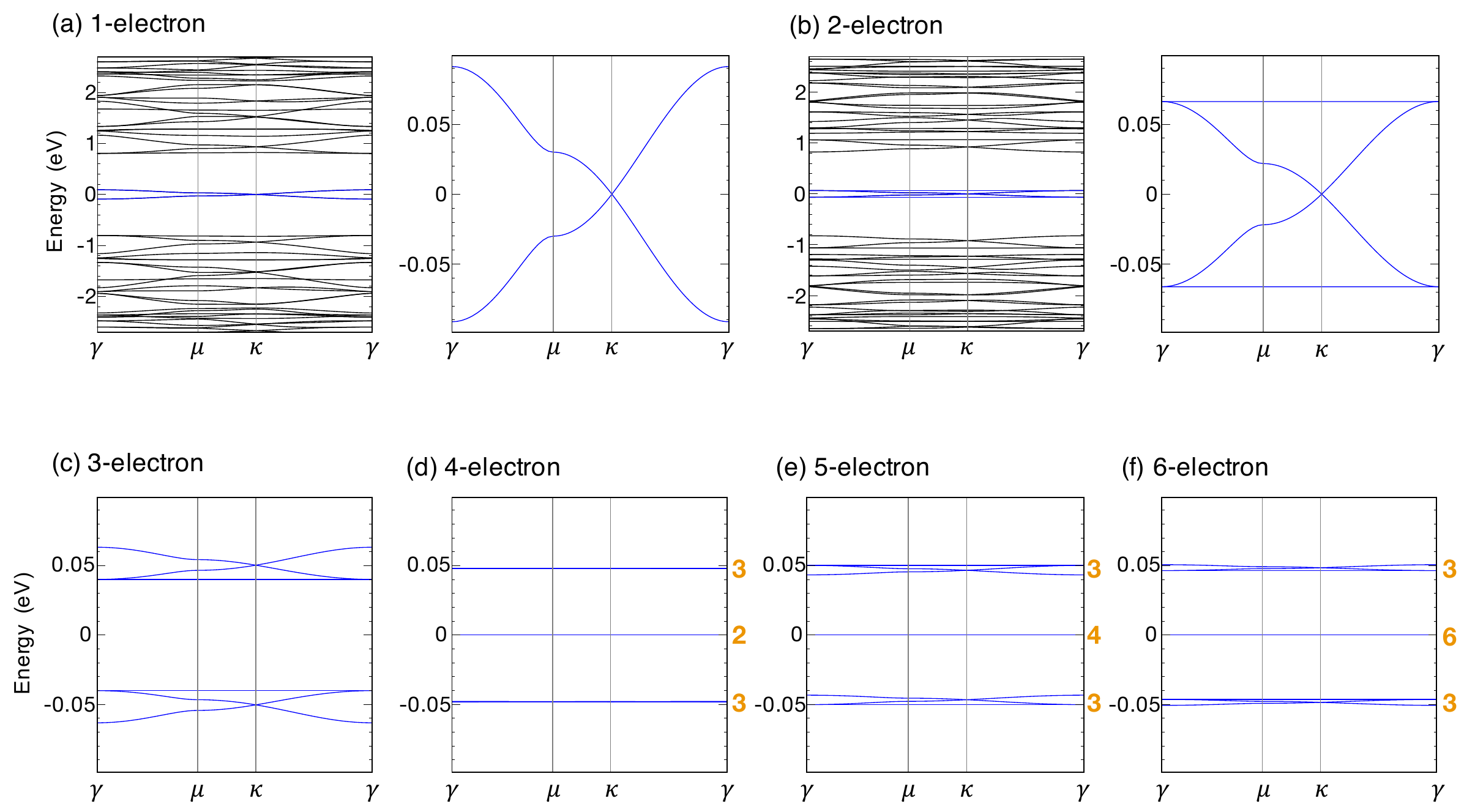}
  \caption{
  Band structures of 2D GNR networks composed of the 1-electron to 6-electron junctions [Fig.\ \ref{fig_GNR_3-junctions}(a) to \ref{fig_GNR_3-junctions}(f)].
In panels of (a) 1-electron and (b) 2-electron junctions, the left and right figures plot the same energy bands in 
wide and narrow energy regions. Orange numbers in (d), (e) and (f) indicate the number of bands (per spin).
  }
 \label{fig_GNR_network_band}
\end{figure*}

We consider honeycomb GNR networks connected by the 1-electron to 6-electron 
junctions [Fig.\ \ref{fig_GNR_3-junctions}(a) to \ref{fig_GNR_3-junctions}(f)].
Here we change only the junction parts (shaded triangles in Fig.\ \ref{fig_GNR_network}), while
we fix the ribbon parts to the $M=7$ ribbon of 4 unit cell long.
Note that the signs of $\Delta N_{\rm tot}$ of junctions at $\alpha$ and $\beta$ are opposite,
so that the sublattice difference is zero in the whole system.
Figure \ref{fig_GNR_network_band} shows the band structures from $\Delta N_{\rm tot} = 1$ to 6.
We see that the topological bound states forms a cluster of energy bands inside the semiconducting energy gap of the GNR.
For the 1-electron case [Fig.\ \ref{fig_GNR_network_band}(a)], we have energy bands analog to graphene.
This is because the only bound state at a single junction belongs to $C_3=1$, 
and it works like a carbon $p_z$ orbital in graphene.
Since the wave function of a single junction is well localized as shown in Fig.\ \ref{fig_GNR_3-junctions}(a),
the system is very well approximated by an effective tight binding model with the nearest neighbor hopping $t$
between the topological bound states. The energy bands are then
given by $E(\mathbf{k}) = \pm t |\sum_{l=1}^3 e^{i \mathbf{k}\cdot\mbox{\boldmath \scriptsize $\lambda$}_l}|$,
where $t \approx 30$ meV in Fig.\ \ref{fig_GNR_network_band}(a).

\begin{figure}
 \centering
 \includegraphics[width=60mm]{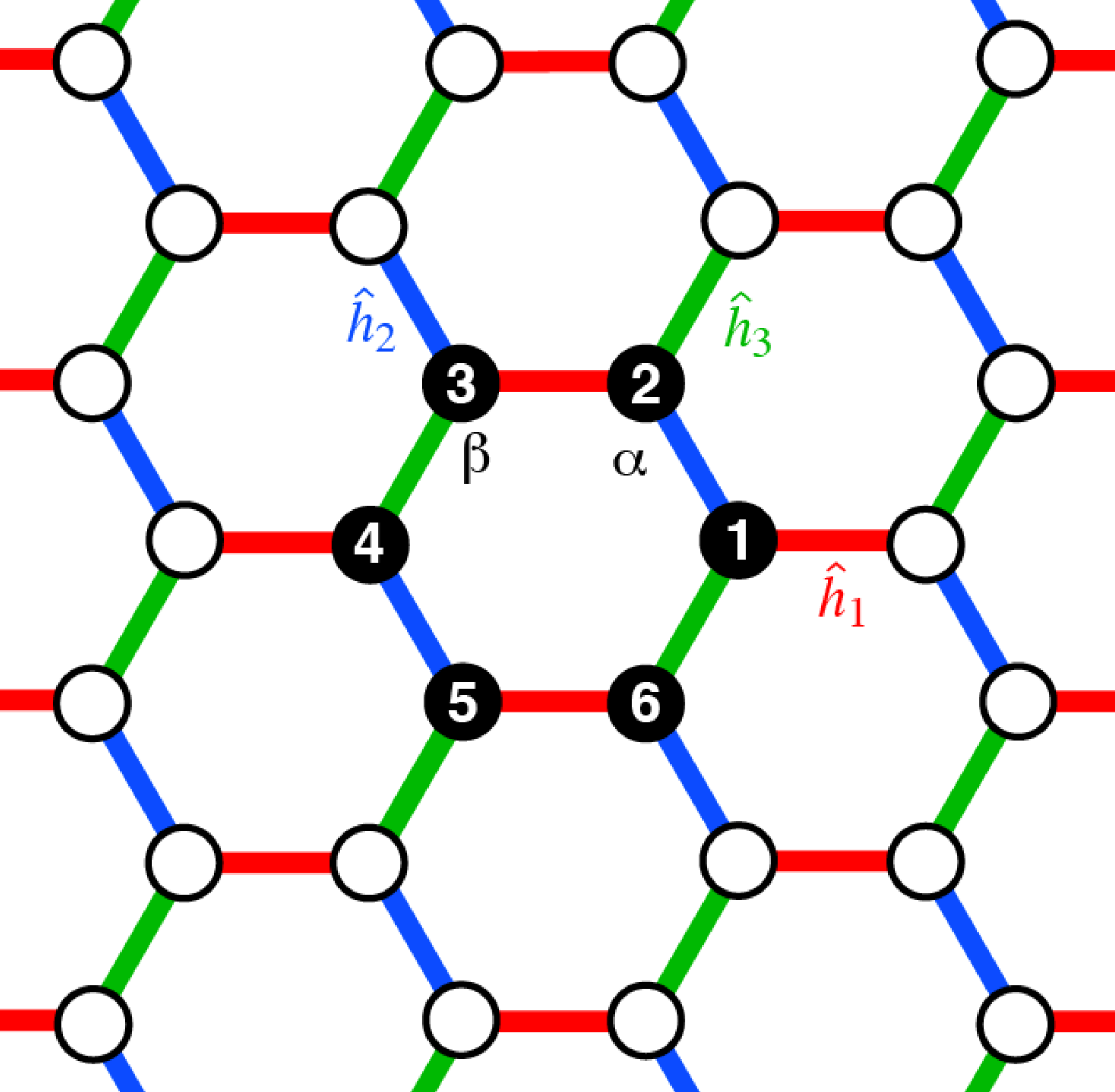}
  \caption{Schematic picture of the effective model for the 2-electron lattice, Eq.\ (\ref{eq_TB}).
 Numbered circles indicate the localized ring state 
 which is an eigenstate of the Hamiltonian (see the text). 
  }
 \label{fig_ring_state}
\end{figure}

For the 2-electron case [Fig.\ \ref{fig_GNR_network_band}(b)], 
we have an ``hour-glass'' band structure composed of graphene-like
bands and a pair of flat bands at the top and bottom.
This is also modeled by an effective tight-binding model based on the symmetry analysis.
According to Table \ref{table1}, we have two zero-energy bound states of $C_3=\omega$ and $\omega^*$,
which are denoted as orbital $1$ and $2$.
If we consider only the nearest neighbor hoppings, the Schr\"{o}dinger equation can be written as
\begin{align}
E \Vec{\Psi}_\beta ({\bf r}) = \sum_{l=1}^3 \hat{h}_l \Vec{\Psi}_\alpha ({\bf r}+\GVec{\lambda}_l),
\label{eq_TB}
\end{align}
where
\begin{align}
&
\hat{h}_1 = 
\begin{pmatrix}
t  & -t'\\
-t' & t
\end{pmatrix}, 
\quad
\hat{h}_2 = 
\begin{pmatrix}
t  & -t'\omega^* \\
-t'\omega & t
\end{pmatrix},
\nonumber\\
&
\hat{h}_3 = 
\begin{pmatrix}
t  & -t'\omega \\
-t'\omega^* & t
\end{pmatrix}.
\end{align}
Here $\Vec{\Psi}_X({\bf r}) = (\Psi^1_X({\bf r}), \Psi^2_X({\bf r}))$ is the wave amplitude at orbital 1 and 2, respectively,  
at the position ${\bf r}$ of the sublattice $X(=\alpha,\beta)$,
$t$ is the hopping amplitude between the same orbitals (1 and 1, or 2 and 2)
and $t'$ is that between different orbitals (1 and 2).
We can show that $t$ and $t'$ must be real due to the time reversal symmetry and $C_2$ symmetry
with respect to the center axis of the ribbon.
The effective lattice model is schematically illustrated in Fig.\ \ref{fig_ring_state}.
The Bloch Hamiltonian in the basis of
$(\Psi^1_\alpha, \Psi^2_\alpha,\Psi^1_\beta,\Psi^2_\beta)$ is 
\begin{align}
 H({\bf k})=
 \mqty(0&h^\dagger({\bf k})  \\ h({\bf k})&0),
\quad
h({\bf k}) =  \sum_{l=1}^3 \hat{h}_l \,  e^{i \mathbf{k}\cdot\mbox{\boldmath \scriptsize $\lambda$}_l}.
\label{eq_H_2-electron_network}
\end{align}

Actually, $t$ and $t'$ are almost equal in the system,
because the hopping integral between the localized states
is dominated by wave overlap in the ribbon part,
where the wave functions of orbital 1 and 2 are identical except for the overall phase factor.
When $t=t'$, the eigen energies of Eq.\ (\ref{eq_H_2-electron_network})
becomes 
\begin{align}
E({\bf k}) = \pm 3t, \,\, \pm t\, \Bigl|\sum_{l=1}^3 e^{i \mathbf{k}\cdot\mbox{\boldmath \scriptsize $\lambda$}_l} \Bigr|,
\end{align}
which are flat bands and graphene-like bands, respectively.  
We have $t\approx t' \approx$ 22 meV in Fig.\ \ref{fig_GNR_network_band}(b).

The emergence of the flat bands is closely related to the existence of a localized eigenstate
analog to a ring state in the kagome lattice.\cite{bergman2008band} 
In the present case, the eigen wave function is given by six spinors on site 1 to 6 in Fig.\ \ref{fig_ring_state},
\begin{align}
&
\Vec{\Psi}(1) = 
\begin{pmatrix}
1 \\ 1
\end{pmatrix},\,
\Vec{\Psi}(2) = 
\pm \begin{pmatrix}
\omega^* \\ \omega
\end{pmatrix},\,
\Vec{\Psi}(3) = 
\begin{pmatrix}
\omega \\ \omega^*
\end{pmatrix},\,
\nonumber\\
&
\Vec{\Psi}(4) = 
\pm \begin{pmatrix}
1 \\ 1
\end{pmatrix},\,
\Vec{\Psi}(5) = 
 \begin{pmatrix}
\omega^* \\ \omega
\end{pmatrix},
\Vec{\Psi}(6) = 
\pm \begin{pmatrix}
\omega \\ \omega^*
\end{pmatrix},\,
\label{eq_ring_state}
\end{align}
where $\pm$ correspond to the eigen energy $E=\pm 3t$.
This is an eigenstate of the Hamiltonian with $t=t'$,
because the six spinors vanish on the operation of the out-going transfer matrices,
i.e.  $\hat{h}_1 \Vec{\Psi}(1) = \hat{h}_3 \Vec{\Psi}(2)  =  \hat{h}_2 \Vec{\Psi}(3) 
= \hat{h}_1 \Vec{\Psi}(4) = \hat{h}_3 \Vec{\Psi}(5) =  \hat{h}_2 \Vec{\Psi}(6) = 0$,
and therefore the state never spreads out to the outer sites when the Hamiltonian is operated.

\begin{figure}
 \centering
 \includegraphics[width=90mm]{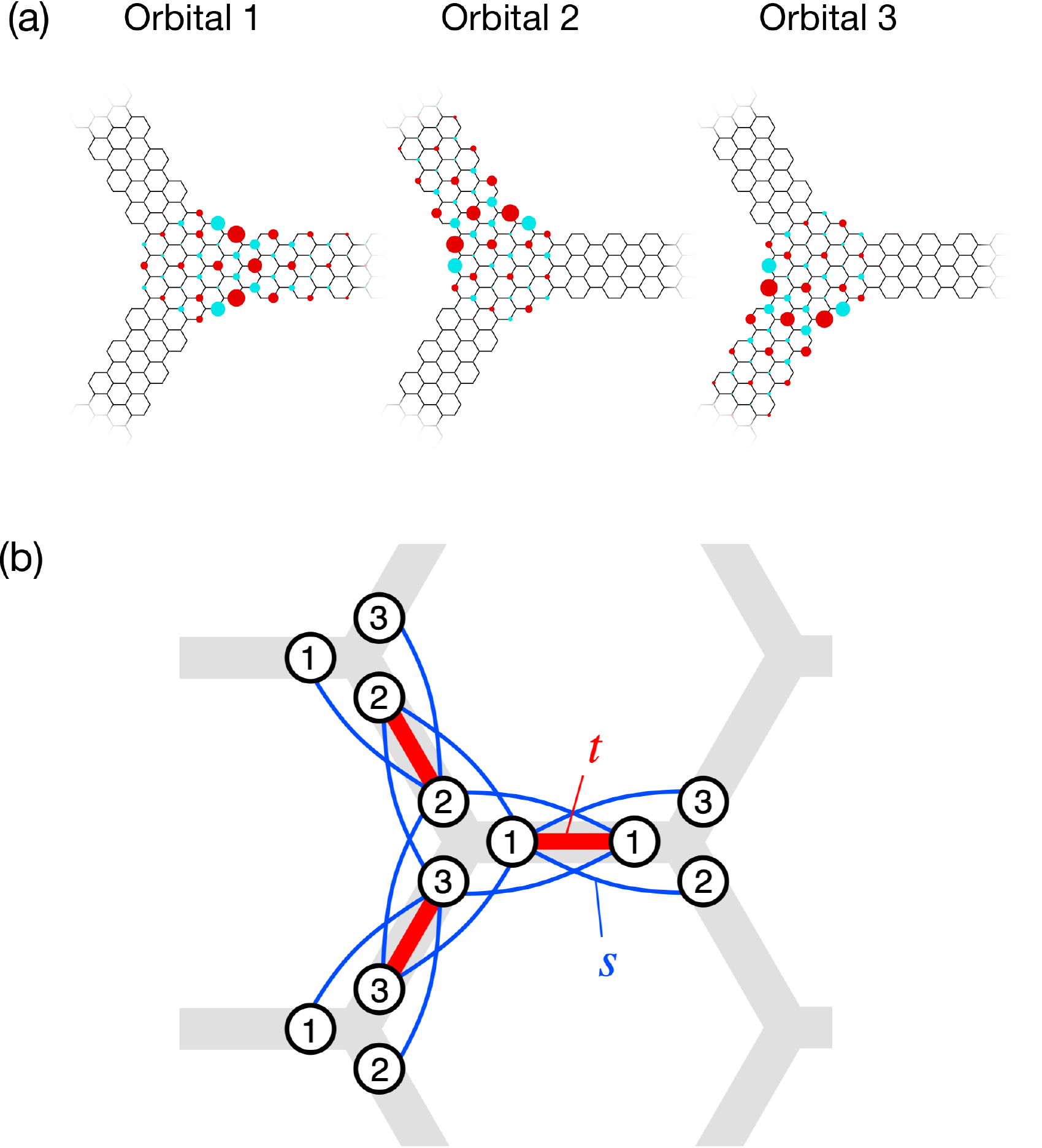}
  \caption{
(a) Three orbitals at a 3-electron junction, obtained by taking linear combinations of the $C_3$ eigenstates in Fig.\ \ref{fig_GNR_3-junctions}.
(b) Corresponding effective lattice model.
  }
 \label{fig_3-electron_network}
\end{figure}

The energy band of the 3-electron network [Fig.\ \ref{fig_GNR_network_band}(c)]
is composed of a pair of kagome-type three-band clusters in upper and lower energies,
which are also explained by a tight-binding model.
In this case, each junction accommodates a set of zero-energy bound states of $C_3=1,\omega$ and $\omega^*$
according to the Table \ref{table1}.
By taking an appropriate linear combination, 
those three states can be rearranged to orbitals 1, 2 and 3,
which are related by 120$^\circ$ rotation as shown in Fig.\ \ref{fig_3-electron_network} (a).
In a honeycomb network, the tight-binding Schr\"{o}dinger equation for those three orbitals can be written 
in the form of Eq.\ (\ref{eq_TB}),
where $\Vec{\Psi}_X({\bf r}) = (\Psi^1_X({\bf r}), \Psi^2_X({\bf r}),\Psi^3_X({\bf r}))\, (X=\alpha,\beta)$ is the 
wave amplitude at orbital 1, 2 and 3, respectively,  and $\hat{h}_l\,(l=1,2,3)$ are $3\times 3$ hopping matrices,
\begin{align}
&
\hat{h}_1 = 
\begin{pmatrix}
t & s & s\\
s & 0 & 0\\ 
s & 0 & 0
\end{pmatrix}, 
\,
\hat{h}_2 = 
\begin{pmatrix}
0  & s & 0\\
s & t & s\\ 
0 & s & 0
\end{pmatrix},
\,
\hat{h}_3 = 
\begin{pmatrix}
0 & 0 & s\\
0 & 0 & s\\ 
s & s & t
\end{pmatrix}.
\end{align}
Here $t$ is the hopping parameter between the nearest neighboring pair,
and $s$ is for the second nearest pair of orbitals, as defined in Fig.\  \ref{fig_3-electron_network}(b). 
Other hopping parameters are tiny and negligible.
The effective lattice structure can be viewed as a tight-binding realization of the Archimedean lattice $(3,12^2)$ \cite{de2019topological},
while the nearest neighbor hopping among the orbital 1, 2 and 3 is absent in the present case.
The Bloch Hamiltonian is again written in the form of Eq.\ (\ref{eq_H_2-electron_network}), and its eigen energies are given by
\begin{align}
E({\bf k})  = \pm (t - 2s),\,\,
\pm \Bigr[t+s \pm s \Bigr|\sum_{l=1}^3 e^{i \mathbf{k}\cdot\mbox{\boldmath \scriptsize $\lambda$}_l}\Bigr| \Bigl].
\end{align}
The band structure in Fig.\ \ref{fig_GNR_network_band}(c) 
is given by $t\approx 48$ meV and $s\approx 3.9$ meV.
The emergence of a pair of kagome bands can be interpreted as follows.
By considering only the strongest coupling $t$, we have the bonding and antibonding states centered at 
the midpoint between neighboring junctions.
Those states are weakly coupled by $s$. Since the midpoints between neighboring sites in the honeycomb lattice form a kagome lattice,
we have a pair of kagome bands from the bonding states and the antibonding states.

The same argument applies to the case of four or more electron junctions.
For the 4-electron network [Fig.\ \ref{fig_GNR_network_band}(d)],
a single junction has a triplet of zero-energy bound states of $C_3=1$, $\omega$ and $\omega^*$,
and also $C_3=1$ completely-localized states as argued.
The triplet gives a pair of kagome lattices just as in the 3-electron network,
but here the parameter $s$ happens to be very small due to an accidental phase cancellation, resulting in almost flat bands at $\pm t$.
The remaining $C_3=1$ state generally gives the graphene-like band as in the 1-electron network,
but in the present model, we have degenerate flat bands at $E=0$.
This is because the corresponding zero-energy junction state is a completely-localized state argued in the previous section,
so its hopping amplitude is exactly zero.
The situation is similar in the 5- and 6-electron networks [Fig.\ \ref{fig_GNR_network_band}(e) and \ref{fig_GNR_network_band}(f)],
where a triplet forms a pair of kagome lattices, while the rest of the completely-localized states
gives a bunch of zero-energy flat bands.

\section{Discussion}
\label{sec_disc}

In realistic graphene samples, the chiral symmetry is broken by the additional effects neglected here,
such as the second nearest neighbor hopping (from A-site to A-site, B-site to B-site;
of the order of 0.1 eV\cite{peres2006electronic}).
The topological junction states are influenced by those extra terms, but they would survive
around zero energy as long as these additional terms are smaller than the semiconducting energy gap of GNR
($\sim$ 1 eV for the $M=7$ ribbon).
For the GNR networks considered in Sec.\ \ref{sec_2D_network}, 
the effective nearest-neighbor tight-binding model for the topological bound states
is expected to be valid even in the presence of the chiral-symmetry breaking terms,
because it is based on the $C_3$ symmetry (the real symmetry of the system)
and also the vanishing of far hopping parameters in the effective lattice
is guaranteed by the localizing feature of the topological bound states.
Including the additional effects neglected in the ideal model 
would shift the existing parameters in the effective tight-binding model, 
while they are not expected to change the model qualitatively.
It should be noted that the completely-localized state found  in Fig.\ \ref{fig_GNR_2-junctions}(d) to \ref{fig_GNR_2-junctions}(f)
is a special feature of the nearest-neighbor tight-binding models of graphene, 
so the resulting full-flat zero-energy bands in Fig.\ \ref{fig_GNR_network_band}(d) to \ref{fig_GNR_network_band}(f)
should have some energy dispersion when the further hoppings are included in graphene.
The realistic band calculation using the density functional approach is left for future research.

The existence of the isolated topological bands inside the bulk energy gap is a major characteristic of our GNR network systems,
which are not seen in graphene superlattices with nanoscale holes in the previous studies.
\cite{yu2008collective,pedersen2008graphene,liu2009band,sinitskii2010patterning,bai2010graphene,liang2010formation,cui2011magic,gunst2011thermoelectric,yang2011inducing,petersen2011clar,oswald2012energy,dvorak2013bandgap,trolle2013large,power2014electronic,chen2018nanoperforated}
One exception is the phenalenyl-phenyl honeycomb network proposed in a recent study \cite{maruyama2016coexistence},
where an isolated narrow band was predicted using the density functional band calculation.
In our language, this is interpreted as a special case of the topological-bound state crystal,
where the central island (phenalenyl) with $\Delta N_{\rm center} =  1$,
and the $M=3$ armchair ribbon (phenylene) with $\Delta N =  0$ give the total topological number $\Delta N_{\rm tot} =  1$.
Indeed, the predicted band structure exhibits a graphene-like band near the zero energy \cite{maruyama2016coexistence},
which is consistent with our general argument.
This example suggests that the present theoretical framework based on the nearest-neighbor tight-binding model
is qualitatively valid in realistic systems.

The topological-state bands in the 2D GNR networks are expected to be an ideal platform 
to emulate a strongly-interacting fermion system on various type of lattices.
Considering well-isolated energy bands and well-localized effective orbitals,
the electron-electron interaction in the topological-state bands
can be incorporated by the on-site Hubbard $U$ in the effective lattice model.
In the junctions in Fig.\ \ref{fig_GNR_3-junctions}, for example, the spread of the wave function is about $r \sim 1$ nm,
where the on-site Coulomb interaction becomes $e^2/(4\pi\epsilon_0 r) \sim 1.4$ eV.
On the other hand, the typical band width $t$ is just 0.1 eV or less,
suggesting that the system is in the strong coupling limit $U/t \gg 1$.
It should also be noted that the effective hopping $t$ can be tuned by changing the ribbon length between the junctions.
Since the topological bound state exponentially decays in space, 
the $t$ exponentially decreases in longer ribbons, pushing the system to even stronger coupling side, and vice versa.

In the literature, a number of theoretical studies have been made on the Hubbard-type models on the honeycomb lattice,
where various types of exotic quantum phases were proposed.\cite{lee2005u,hermele20072,uchoa2007superconducting,raghu2008topological,meng2010quantum}.
It was also predicted that the chiral $d$-wave superconductivity 
emerges in the doped honeycomb lattice in the limit of $U/t \gg 1$ \cite{nandkishore2012chiral,black2014chiral},
which is expected to be realized in the current system of $\Delta N_{\rm tot}=1$.
The many body physics in the ``hour-glass'' lattice ($\Delta N_{\rm tot}=2$) and the kagome lattice  ($\Delta N_{\rm tot}=3$) 
would also be intriguing problems, particularly in relation to the flat band physics.

Lastly, while we concentrated on honeycomb GNR superlattices with varying the number of electrons per junction,
it would be interesting to consider topological metamaterials on other types of networks,
such as triangular, square, kagome, and Archimedean lattices. \cite{de2019topological}

\section{Conclusion}
We studied the topological bound states in general junction structures of chiral-symmetric systems.
The general formulation developed in this work allows us to estimate the number of topological states 
in any multiway junctions composed of an arbitrary number of channels.
By using the method, we calculate the zero-energy bound states in various types of two-way and three-way junctions
of graphene nanoribbons.
In the latter part, we calculated the energy bands of 2D armchair GNR networks,
and demonstrated that the topological junction states 
form an ideal tight-binding bands energetically isolated from the rest of the spectrum.
The $Z$ number of a single junction determines the number of orbitals in a single effective atom,
and depending on it, we have different types of nanoscale effective materials.
We expect that the system serves as a quantum simulator of the Hubbard like model in the strong coupling regime,
and it would provide an ideal platform to emulate interacting fermion systems.
\label{sec_concl}

\begin{acknowledgements}
This work was supported by JSPS KAKENHI Grant No. JP17K05496, No. JP20H01840, 
JP20H00127, No. JP16K17755, and No. JP20K14415.
\end{acknowledgements}

\bibliography{GNR_tamaki}

\end{document}